# Production of complex organic molecules: H-atom addition versus UV irradiation

K.-J. Chuang,[1,2] G. Fedoseev,[1]† D. Qasim,[1] S. Ioppolo,[3] E.F. van Dishoeck[2] and H. Linnartz[1]

[1]*Sackler Laboratory for Astrophysics, Leiden Observatory, Leiden University, PO Box 9513, NL-2300 RA Leiden, the Netherlands*
[2]*Leiden Observatory, Leiden University, PO Box 9513, NL-2300 RA Leiden, the Netherlands*
[3]*School of Physical Sciences, The Open University, Walton Hall, Milton Keynes MK7 6AA, UK*

## ABSTRACT

Complex organic molecules (COMs) have been identified in different environments in star-forming regions. Laboratory studies show that COMs form in the solid state, on icy grains, typically following a 'non-energetic' (atom-addition) or 'energetic' (UV-photon absorption) trigger. So far, such studies have been largely performed for single processes. Here, we present the first work that quantitatively investigates both the relative importance and the cumulative effect of '(non-)energetic' processing. We focus on astronomically relevant $CO:CH_3OH = 4:1$ ice analogues exposed to doses relevant for the collapse stage of dense clouds. Hydrogenation experiments result in the formation of methyl formate (MF; $HC(O)OCH_3$), glycolaldehyde (GA; $HC(O)CH_2OH$) and ethylene glycol (EG; $H_2C(OH)CH_2OH$) at 14 K. The absolute abundances and the abundance fractions are found to be dependent on the H-atom/$CO:CH_3OH$-molecule ratios and on the overall deposition rate. In the case that ices are exposed to UV photons only, several different COMs are found. Typically, the abundance fractions are 0.2 for MF, 0.3 for GA and 0.5 for EG as opposed to the values found in pure hydrogenation experiments without UV in which MF is largely absent: 0.0, 0.2–0.6 and 0.8–0.4, respectively. In experiments where both are applied, overall COM abundances drop to about half of those found in the pure UV irradiation experiments, but the composition fractions are very similar. This implies COM ratios can be used as a diagnostic tool to derive the processing history of an ice. Solid-state branching ratios derived here for GA and EG compare well with observations, while the MF case cannot be explained by solid-state conditions investigated here.

## 1 INTRODUCTION

Complex organic molecules (COMs) are regarded as the building blocks of species that are inherent to life. In space, COMs have been unambiguously identified in very different environments (Herbst & van Dishoeck 2009), such as toward the Galactic Centre in hot cores and giant molecular clouds (Hollis, Lovas & Jewell 2000; Martín-Pintado et al. 2001; Requena-Torres et al. 2006, 2008; Belloche et al. 2013), toward low- and high-mass proto-stars (Blake et al. 1987; Neill et al. 2014; Tercero et al. 2015; Jørgensen et al. 2012, 2016; Caselli & Ceccarelli 2012; Coutens et al. 2015; Taquet et al. 2015; Rivilla et al. 2016), molecular outflows (Arce et al. 2008; Codella et al. 2015), photon-dominated regions (Guzmán et al. 2013), as well as toward dark cloud cores and pre-stellar cores (Marcelino et al. 2007; Bacmann et al. 2012; Öberg et al. 2010, 2011; Cernicharo et al. 2012; Vastel et al. 2014; Jiménez-Serra et al. 2016). With eight and 10 atoms, the simplest sugar, glycolaldehyde (HC(O)CH$_2$OH, hereafter GA), and the simplest sugar alcohol, ethylene glycol (H$_2$C(OH)CH$_2$OH, hereafter EG), can be considered amongst the largest organic molecules detected so far in the interstellar medium (ISM). GA has been observed toward solar-mass proto-stars (Jørgensen et al. 2012, 2016; Coutens et al. 2015) and in comets (Crovisier et al. 2004; Biver et al. 2014; Goesmann et al. 2015; Le Roy et al. 2015) together with chemically related species, such as EG or methyl formate (CH(O)OCH$_3$, hereafter MF), one of the corresponding isomers. It is generally accepted that these species form in the solid state, on icy dust grains and are often found chemically linked to other O-bearing COMs in star-forming regions (Rivilla et al. 2016). Recent work on the spatial distribution of COMs in the L1544 pre-stellar core (with $T < 10$ K) revealed that O-bearing gas-phase COMs are more abundant toward a low-density ($A_v \sim 7.5$–8 mag) shell, where gas-phase methanol also peaks, than toward the denser ($A_v \geq 30$ mag) continuum peak (Jiménez-Serra et al. 2016). At $A_v \sim 7$–8 mag, CO molecules start freezing-out onto the grains (Pontoppidan 2006; Boogert, Gerakines & Whittet 2015). An enhancement of factors $\sim$2–10 in the abundance of O-bearing COMs in the outer shell of L1544 can be explained by the surface formation of COMs in a CO-rich ice environment followed by non-thermal desorption. In the outer shell, visual extinctions are indeed high enough for COMs to avoid photo-dissociation by the external interstellar radiation field, but they are not high enough for COMs to be completely accreted onto the ice grains (Jiménez-Serra et al. 2016). In such environments, both 'non-energetic' (atom bombardment) and 'energetic' (cosmic rays (CR) induced UV photons) formation routes can result in the abundance of COMs.

* E-mail: chuang@strw.leidenuniv.nl
† Present address: INAF–Osservatorio Astrofisico di Catania, via Santa Sofia 78, I-95123 Catania, Italy.

In dense clouds, CO is the second most abundant ice species along with CO$_2$ on grain surfaces with a relative abundance ranging from 20 to 45 per cent with respect to water ice (Boogert, Gerakines & Whittet 2015). Most of the CO ice is formed through accretion from the gas-phase during the so-called catastrophic CO freeze-out stage, when it accretes onto layers of preformed water ice, resulting in a CO-rich apolar ice coating (Tielens et al. 1991; Pontoppidan 2006; Gibb et al. 2004; Öberg et al. 2011; Mathews et al. 2013; Boogert, Gerakines & Whittet 2015). At this stage, CO is the starting point in hydrogenation schemes that result in H$_2$CO and CH$_3$OH formation through successive ('non-energetic') H-atom addition reactions. This surface reaction scheme is generally considered as the pathway to explain the observed gaseous CH$_3$OH abundances in space (Geppert et al. 2005), a conclusion that is supported by theoretical and modelling studies (Tielens & Hagen 1982; Shalabiea & Greenberg 1994; Cuppen et al. 2009; Vasyunin & Herbst 2012) as well as compelling laboratory experiments (Hiraoka et al. 1994; Watanabe & Kouchi 2002; Zhitnikov & Dmitriev 2002; Fuchs et al. 2009; Linnartz, Ioppolo & Fedoseev 2015). CH$_3$OH ice has been detected with abundances ranging from 1 to 25 per cent with respect to water ice (Bottinelli et al. 2010; Boogert et al. 2015). Indeed, spectroscopic studies of CO ice, comparing astronomical spectra (Pontoppidan et al. 2003; Penteado et al. 2015) and laboratory spectroscopic ice data (LDI[1]), confirm that CH$_3$OH and CO are intimately mixed in interstellar ices (Cuppen et al. 2011; Penteado et al. 2015). In the follow-up studies (Fedoseev et al. 2015; Butscher et al. 2015; Chuang et al. 2016), it was shown that these 'non-energetic' reactions also offer pathways toward COM formation at temperatures as low as 10 K along steps of the CO $\xrightarrow{2H}$ H$_2$CO $\xrightarrow{2H}$ CH$_3$OH reaction scheme, through recombination reactions of reactive intermediates, e.g. HCO, CH$_2$OH and CH$_3$O, that form both in addition and abstraction reactions. This results in the unambiguous formation of GA, EG and MF, fully in line with astrochemical models that hint at COM formation early in the cosmochemical cycle, namely in dark interstellar clouds (Charnley & Rodgers 2005; Woods et al. 2012).

Laboratory studies involving 'energetic' processing of CH$_3$OH ice, e.g. UV photolysis (Gerakines, Schutte & Ehrenfreund 1996; Öberg et al. 2009; Paardekooper, Bossa & Linnartz 2016a; Öberg

2016), electron impact (Henderson & Gudipati 2015; Maity, Kaiser & Jones 2015) and ion bombardment (Moore & Hudson 2005; Modica & Palumbo 2010), show that it is also possible to form GA, EG and MF as well as a several other COMs in CH$_3$OH-rich interstellar ice analogues. Upon irradiation, for example, the methanol dissociates, and the reactive free radicals recombine, forming larger species, such as acetaldehyde, ethanol and dimethyl ether, as well as the three COMs that are investigated here.

Clearly, completely different triggers, involving similar reactive intermediates, result in the formation of similar species, but not necessarily with comparable (relative) abundances, and this knowledge may be used as a tool to relate the role of specific chemical processes to specific astronomical environments. In dark and dense clouds, impacting H atoms are the dominating process and thermal processing can be fully neglected. However, the internal UV radiation field, which is due to cosmic-ray excitation of molecular hydrogen, can influence the ice chemistry. In translucent clouds, hot cores, and proto-planetary disks, the 'energetic' processing will be much more relevant; thermal effects are at play for the higher temperatures and particularly UV radiation fields are much more intense than in dense cold quiescent regions. Moreover, further gas-phase chemistry leads to the formation of COMs upon sublimation of second-generation species from icy dust grains (Charnley et al. 1992; see also Herbst & van Dishoeck 2009 for a review). Recently, Vasyunin & Herbst (2013), Balucani, Ceccarelli & Taquet (2015) and Taquet, Wirstrom & Charnley (2016) proposed that MF and other complex species may also be produced through low temperature gas-phase routes. However, the abundances derived from their models do not always match the observational data. Finally, COMs have also been identified mass spectrometrically in cometary objects (Goesmann et al. 2015; Le Roy et al. 2015) that consist of the planetesimals built during the proto-planetary stage. The investigation of the chemical connection between interstellar and cometary ices is an important future goal (Mumma & Charnley 2011) that has become further within reach through all recent data for 67P and the launch of JWST in 2018.

In this paper, we present for the first time laboratory studies that allow to *quantitatively* compare COM formation through hydrogenation and UV-induced reactions as well as their cumulative effect in interstellar ice analogues. The combined experiments are relevant to conditions in dark interstellar clouds as described above and have been applied in the past for other systems. Watanabe et al. (2007) investigated the competition between hydrogenation and photolysis in a pre-deposited H$_2$O:CO ice mixture. Another more recent study showed that, for the case of NO hydrogenation and UV photolysis, hydroxylamine (NH$_2$OH) is efficiently formed upon hydrogenation of NO molecules, but once UV photons are added, the presence of NH$_2$OH in the ice diminishes and the available nitrogen gets locked in three chemically linked species: HNCO, OCN$^-$ and H$_2$NCHO (Congiu et al. 2012; Fedoseev et al. 2016). This finding has been taken as one of the arguments for the non-detection of NH$_2$OH in space despite a number of dedicated surveys and to explain the observed correlation between HNCO and H$_2$NCHO abundances (Pulliam, McGuire & Remijan 2012; McGuire et al. 2015).

Here, we focus on three specific COMs: GA, EG and MF. As stated, these three COMs are amongst the most commonly produced complex species in both 'energetic' and 'non-energetic' ice processing. We compare the laboratory findings for the MF/EG, MF/GA and GA/EG solid-state abundance ratios for different experiments with those found for the gas-phase abundance ratios in different astronomical environments. In this approach, we make two assumptions: first, we consider EG, GA and MF as representative products in the COM formation schemes, realizing that we neglect other (higher generation) COMs that form, but typically with lower abundances; second, it is not a priori clear that solid-state and gas-phase abundance ratios scale one to one. This depends on the sublimation process that bridges the grain–gas gap and that may be triggered by shocks, heat or chemical, reactive and photo-desorption processes (Garrod, Weaver & Herbst 2008; Herbst & van Dishoeck 2009; Öberg et al. 2009; Fayolle et al. 2011; Minissale et al. 2016). Moreover, the sublimated species could subsequently participate in further chemical reactions in the gas phase. Whereas thermal desorption is a smooth and fast process leaving the ice molecules intact, photo-desorption tends to dissociate COMs upon excitation. This will be discussed in the final section. First the experimental procedure is described and subsequently the results are discussed.

---

[1] See the Leiden Database for Ice at http://icedb.strw.leidenuniv.nl/, which is maintained by the Sackler Laboratory for Astrophysics.

## 2 EXPERIMENTAL PROCEDURE

### 2.1 Experimental set-up

All experiments are performed under ultra-high vacuum (UHV) conditions, using SURFRESIDE2. Details of this set-up and measurement procedures have been described by Ioppolo et al. (2013). The experiments are performed in a main chamber with a base pressure of ~10⁻¹⁰ mbar. In its centre, a gold-coated copper substrate is positioned so that it is connected to a closed-cycle helium cryostat, reaching temperatures as low as 13 K and onto which ices are grown. Resistive heating allows temperatures up to 330 K. In order to perform hydrogenation and UV irradiation experiments simultaneously, the icy substrate can be exposed to H atoms formed in a Hydrogen Atom Beam Source (HABS; Tschersich 2000), mounted in a second UHV chamber, and UV photons generated by a Microwave Discharge Hydrogen flowing Lamp (MDHL) that is directly attached to the main chamber (see Fedoseev et al. 2016 for more details). The UV light is guided through an MgF₂ window and hits the substrate at normal incidence, covering the whole area of the gold substrate (2.5 × 2.5 cm²). The optical axis of the lamp intersects the surface at a 45° angle with the H-atom beam and at a 22° angle with the two available molecular deposition lines. The H-atom flux of the HABS and UV-photon flux as well as the spectrum of the MDHL have been well calibrated in a number of previous experiments (see Ioppolo et al. 2013; Ligterink et al. 2015). The UV spectrum ranging between 114 and 220 nm consists of a main peak at 121.6 nm (Ly-α, 33 per cent of the total flux) and a series of other peaks between 155 and 165 nm (H₂ emission, 20 per cent of the total flux) superposed on a broad continuum (47 per cent of the total flux). The total UV flux is estimated to be $4.1 \pm 0.5 \times 10^{12}$ photons cm⁻²s⁻¹, with ~13 per cent of relative uncertainty. The laboratory UV spectrum resembles quite accurately the spectrum produced by CRs interacting with H₂ in dense clouds. H-atom fluxes applied here are $6 \times 10^{12}$ and $1 \times 10^{12}$ atoms cm⁻² s⁻¹ in high and low co-deposition experiments, respectively. In dense clouds, icy dust grains experience both atom addition and UV irradiation. This results in a 3:2 H-atom to UV-photon exposure; the H-atom flux (~10⁴ atoms cm⁻² s⁻¹) is either comparable or higher than that of the UV photons (CR-induced UV-photon flux 1–10 × 10³ photons cm⁻² s⁻¹; see Prasad & Tarafdar 1983; Mennella et al. 2003; Shen et al. 2004). It should be noted that an absolute H-atom flux is given. The effective H-atom flux can be significantly lower. This is due to the difference in sticking probability between room temperature H atoms as used in the laboratory and 10–20 K cold H atoms as expected in dense clouds. Furthermore, the ~10 orders of magnitude higher H-atom flux used in the laboratory results in a significantly higher H-atom recombination rate, reducing the amount of H atoms available for the reaction with CO.

### 2.2 Experimental methods

All performed experiments utilize a co-deposition technique, which is the simultaneous deposition of ices with H atoms and/or UV photons. This overcomes the problem of a limited H-atom penetration depth into the bulk of the ice compared to the larger UV-photon penetration depth and, also, is more representative of the actual processes taking place in the ISM, where atoms and molecules continuously adsorb onto grains. The chosen CO:CH₃OH = 4:1 ratio is a representative value for ice ratios found in quiescent dense clouds observations toward background stars and observations of low-mass young stellar objects (YSOs) (see Boogert et al. 2015), although it should be stressed that other mixing ratios have been observed as well (Pontoppidan, van Dishoeck & Dartois 2004; Cuppen et al. 2011). The CO:CH₃OH ice mixtures are grown by simultaneous deposition of CO and CH₃OH on top of a pre-deposited (10 Langmuir) argon (Ar) ice. The Ar layer further prevents direct interactions between deposited molecules and photo-electrons potentially produced by the interaction between the incident UV photons and the gold substrate (gold work function >4.7 eV; see Hopkins & Riviere 1964). CO (Linde 2.0) is used for ice deposition and H₂ (Praxair 5.0) is used for the H-atom source and MDHL. Liquid CH₃OH (Sigma-Aldrich, 99.8 per cent) is purified through three freeze–pump–thaw cycles before it is used as a mixture component. The ice thickness is derived using a modified Beer's law with a calibrated absorbance strength that is measured by He–Ne laser interference experiments during ice growth on SURFRESIDE2, following a similar procedure as described by Paardekooper et al. (2016b).

Ices are monitored *in situ* by Fourier Transform Reflection-Absorption InfraRed Spectroscopy (FT-RAIRS) in the range from 700 to 4000 cm⁻¹, with 1 cm⁻¹ resolution. After completion of a co-deposition experiment, a temperature-programmed desorption experiment using a Quadrupole Mass Spectrometer (TPD QMS) is performed. For each species, the desorption temperature and ionized fragment pattern are unique and used to identify the resulting COMs. In order to compare GA, EG and MF yields, the desorption QMS signals are integrated, then normalized to the total column density of the most abundant carbon-bearing species observed by RAIRS at the end of the co-deposition (see Fedoseev et al. 2015; Chuang et al. 2016). Then, the obtained values for GA, EG and MF are calibrated using the available literature ionisation cross-section values (Hudson et al. 2003, 2006; Bull & Harland 2008). The uncertainty in the abundance determination is statistically derived by averaging the result from a number of identical experiments. The lack of HCO, CH₃O and CH₂OH observations at the end of the hydrogenation experiments (Chuang et al. 2016) shows that the radical recombination during the heating process is negligible under our experimental conditions. Table 1 lists the relevant experiments performed for this work.

## 3 RESULTS AND DISCUSSION

Experiments 1–3 are performed to study and compare the isolated effect of hydrogenation and UV photolysis and the combined effect of both, respectively, on the formation of COMs in a co-deposited CO:CH₃OH = 4:1 mixture at 14 K. All three experiments are intentionally performed under the same experimental conditions and for well-chosen parameter settings in order to compare the different formation/destruction mechanisms. Also control experiments are performed, directly linked to exps. 1 (1.1–1.3) and 3 (3.1–3.3) or for other settings (4.1–4.3). The details are summarized in Table 1.

**Table 1.** Overview of the performed experiments.

| No. | Experiments | $T_{sample}$ (K) | $Flux_{(CO+CH_3OH)}$ (cm$^{-2}$ s$^{-1}$) | Ratio (CO:CH$_3$OH) | $Flux_{XH}$ (cm$^{-2}$ s$^{-1}$) | $Flux_{XUV}$ (cm$^{-2}$ s$^{-1}$) | Time (s) | COM composition fraction MF HC(O)OCH$_3$ | GA HC(O)CH$_2$OH | EG H$_2$C(OH)CH$_2$OH |
|---|---|---|---|---|---|---|---|---|---|---|
| 1 | CO + CH$_3$OH + H | 14 | 1.2E13 | 4:1 | 6.0E12 | — | 3600 | 0.0 | 0.2 | 0.8 |
| 2 | CO + CH$_3$OH + hν | 14 | 1.2E13 | 4:1 | — | 4.0E12 | 3600 | 0.2 | 0.3 | 0.5 |
| 3 | CO + CH$_3$OH + H + hν | 14 | 1.2E13 | 4:1 | 6.0E12 | 4.0E12 | 3600 | 0.2 | 0.3 | 0.5 |

| No. | Control experiments | $T_{sample}$ (K) | $Flux_{(CO+CH_3OH)}$ (cm$^{-2}$ s$^{-1}$) | Ratio (CO:CH$_3$OH) | $Flux_{XH}$ (cm$^{-2}$ s$^{-1}$) | $Flux_{XUV}$ (cm$^{-2}$ s$^{-1}$) | Time (s) | COM composition fraction MF HC(O)OCH$_3$ | GA HC(O)CH$_2$OH | EG H$_2$C(OH)CH$_2$OH |
|---|---|---|---|---|---|---|---|---|---|---|
| 1.1 | CO + CH$_3$OH + H$_2$ | 14 | 1.2E13 | 4:1 | — | — | 3600 | — | — | — |
| 1.2 | CO + CH$_3$OH + H | 14 | 2.0E13 | 4:1 | 1.0E12 | — | 21600 | <0.05 | 0.4 | 0.6 |
| 1.3 | CO + CH$_3$OH + H | 14 | 2.0E12 | 4:1 | 6.0E12 | — | 21600 | <0.05 | 0.6 | 0.4 |
| 3.1 | CO + CH$_3$OH + H$_2$ | 14 | 1.2E13 | 4:1 | — | — | 3600 | — | — | — |
| 3.2 | CO + CH$_3$OH + H$_2$ (100%) + hν | 14 | 1.2E13 | 4:1 | — | 4.0E12 | 3600 | 0.2 | 0.3 | 0.5 |
| 3.3 | CO + CH$_3$OH + H$_2$ (70%) + hν | 14 | 1.2E13 | 4:1 | — | 4.0E12 | 3600 | 0.2 | 0.3 | 0.5 |
| 4.1 | CO + CH$_3$OH + Ar(100%) + hν | 14 | 1.2E13 | 4:1 | — | 4.0E12 | 3600 | 0.20 | 0.25 | 0.55 |
| 4.2 | CO + CH$_3$OH + Ar(70%) + hν | 14 | 1.2E13 | 4:1 | — | 4.0E12 | 3600 | 0.2 | 0.3 | 0.5 |
| 4.3 | CO + CH$_3$OH + Ar(30%) + hν | 14 | 1.2E13 | 4:1 | — | 4.0E12 | 3600 | 0.2 | 0.3 | 0.5 |

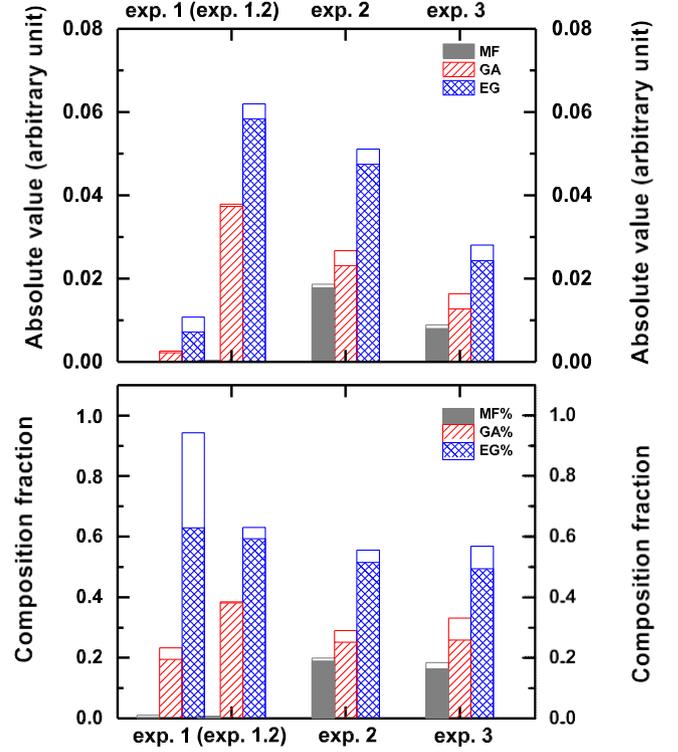

**Figure 1.** Absolute abundances (upper panel) and composition fractions (lower panel) for MF, GA and EG as obtained in QMS TPD experiments for pure hydrogenation (exps. 1 and 1.2), pure UV photolysis (exp. 2) and simultaneous UV-photon irradiation and H-atom addition (exp. 3), all for 14 K CO:CH$_3$OH = 4:1 ice mixtures. Experiment 1.2 is performed under the same experimental conditions and overall fluence as exp. 1, except its timescale is six times longer. Values are calibrated using the corresponding ionization cross-sections and are normalized for the total amount of carbon-bearing molecules observed before performing the TPD experiment. White bars show the standard errors estimated from independent experiments.

The results are summarized in Fig. 1 that shows the absolute abundances (upper panel) and composition fractions (lower panel) for MF, GA and EG resulting from experiments 1–3. A comparison of the absolute abundances for the three experiments reveals that under the selected and identical experimental conditions (i) H-atom addition induced surface reactions lead to the lowest values for the three COMs. It should be noted that to select an astrophysical relevant H-atom/UV-photon flux of ∼1, a high H-atom flux is used for this set of experiment. However, by performing an experiment that has the same overall H-atom fluence of exps. 1 and 3 in the longer timescale (i.e. lower flux and longer exposition time; exp. 1.2), the COM yields resemble those of the pure UV-photolysis experiment except for MF. A lower H-atom flux is closer to the conditions in the ISM. These flux-dependent experiments are discussed in Section 3.1.1; (ii) the highest amount of COMs comes from pure UV photolysis of the ice; (iii) the combination of hydrogenation and UV photons leads to lower amounts of COMs than in the case of pure UV photolysis. These findings are briefly introduced and discussed in more detail from Section 3.1 onward.

In exp. 1, smaller absolute amounts of GA and EG are detected from TPD QMS data and the composition fraction of the three

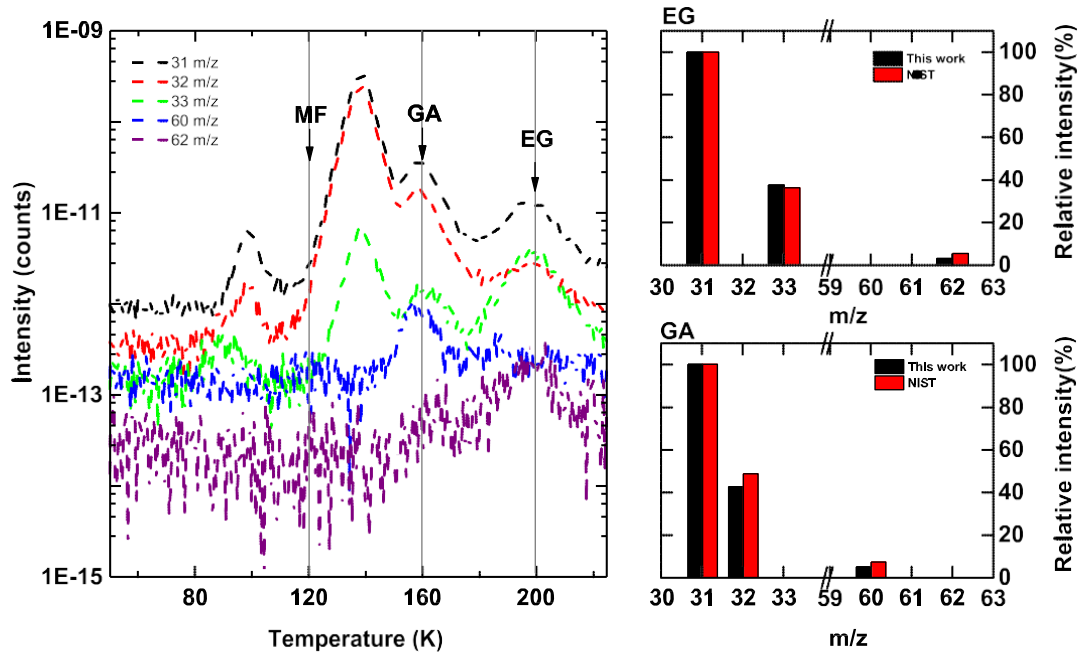

**Figure 2.** The left-hand panel shows part of TPD QMS spectra obtained after co-deposition of CO:CH$_3$OH = 4:1 ice mixtures with H-atom flux $6.0 \times 10^{12}$ atoms cm$^{-2}$s$^{-1}$ at 14 K (exp. 1.3) for $m/z$ = 31, 32, 33, 60, 62 amu. Peaks at $T \sim$ 120, 160 and 200 K correspond to MF, GA and EG. For the latter two species, their fragmentation patterns upon 70 eV electron impact ionization are compared to the available literature values on the right-hand panels.

COMs is given as 0.0 for MF, 0.2 for GA and 0.8 for EG, respectively. In this case, the formation of MF in detectable amounts is only unambiguously confirmed when UV light is used. Indeed, H-atom-induced abstraction from CH$_3$OH results in the preferential formation of CH$_2$OH rather than CH$_3$O radicals (Hidaka et al 2009; Chuang et al. 2016). The CH$_3$O radical can be produced by the hydrogenation of H$_2$CO that, however, is not one of the initial components of the ice mixture and therefore mainly builds-up upon hydrogenation of CO molecules. On the contrary, CH$_3$O is abundantly formed upon photolysis of CH$_3$OH ice. The COM formation yields obtained in exp. 2 are the highest of the three selected experiments and are ~50 per cent higher than those found in exp. 3. This is not surprising because the impinging UV photons can efficiently release their energy to both the surface and bulk of the ice. As a consequence, many molecular bonds are broken and, in timescales of picoseconds, the molecular fragments can recombine giving rise to a rearrangement of the chemical structure that leads to the formation of new and often more complex molecular species. In exp. 3, an overabundance of H atoms can reduce COM formation rates through their recombination with the HCO, CH$_2$OH and CH$_3$O radicals produced by UV photons. This favours the production of simpler species and rather uniformly reduces the formation of more complex ones. As discussed later in the text, segregation by H$_2$ molecules, i.e. the case in which free radicals produced by photo-dissociation are spatially separated by abundantly present H$_2$ molecules, also reduces COM yield rates. It should be noted that although the cumulative effect of 'energetic' and 'non-energetic' processes frustrates COM formation, this should not be generalized. For instance, in Fedoseev et al. (2016), NH$_2$OH is formed by the hydrogenation of NO and simultaneously destroyed by UV photolysis to form HNCO, OCN$^-$ and NH$_2$CHO, when the initial NO ice is mixed in CO/H$_2$CO/CH$_3$OH-rich environments. In the work presented here, the composition fractions of COMs for exps. 2 and 3 are similar and amount to ~0.2 for MF, ~0.3 for GA and ~0.5 for EG. This indicates that the larger amount of COMs still originates from UV-induced reactions even for exp. 3 under our experimental conditions. Below, a detailed discussion of the results from exps. 1–3 is presented, as well as from a series of selected control experiments (exps. 1.1-4.3) performed to constrain the chemistry and physics at play.

### 3.1 Hydrogenation of CO:CH$_3$OH ice mixtures

A set of experiments (exps. 1 and 1.1–1.3) is performed to study the hydrogenation of CO:CH$_3$OH ice mixtures for different overall deposition rates and H/ices ratio settings. Fig. 2 presents the results of a typical experiment (exp. 1.3). Here QMS TPD spectra are shown and obtained after hydrogenation of a CO:CH$_3$OH ice mixture at 14 K. Previous laboratory studies have shown that upon CO ice hydrogenation not only H$_2$CO and CH$_3$OH form (Watanabe & Kouchi 2002; Fuchs et al. 2009), but also GA and EG are produced (Fedoseev et al. 2015). The experiments presented here are fully consistent with this finding. Two of the TPD peaks desorbing at lower temperature are due to H$_2$CO (~100 K) and CH$_3$OH (~140 K). Two other peaks at higher temperatures can be assigned to species with heavier masses. The first peak is located around 160 K, and the second one around 200 K. Based on the available desorption values reported by Öberg et al. (2009), and in recent work by Fedoseev et al. (2015) and Maity, Kaiser & Jones (2015), these two peaks are assigned to GA and EG, respectively. In addition, their QMS spectra for 70 eV electron ionization energy are largely consistent with the corresponding NIST database values (see footnote[2]). Only traces of MF (a peak of 60 $m/z$ centred around 120 K) can be observed in this experiment.

[2] NIST Mass Spec Data Center, S.E. Stein, director, "Mass Spectra" in NIST Chemistry WebBook, NIST Standard Reference Database Number 69, Eds. P.J. Linstrom and W.G. Mallard, National Institute of Standards and Technology, Gaithersburg MD, 20899, http://webbook.nist.gov.

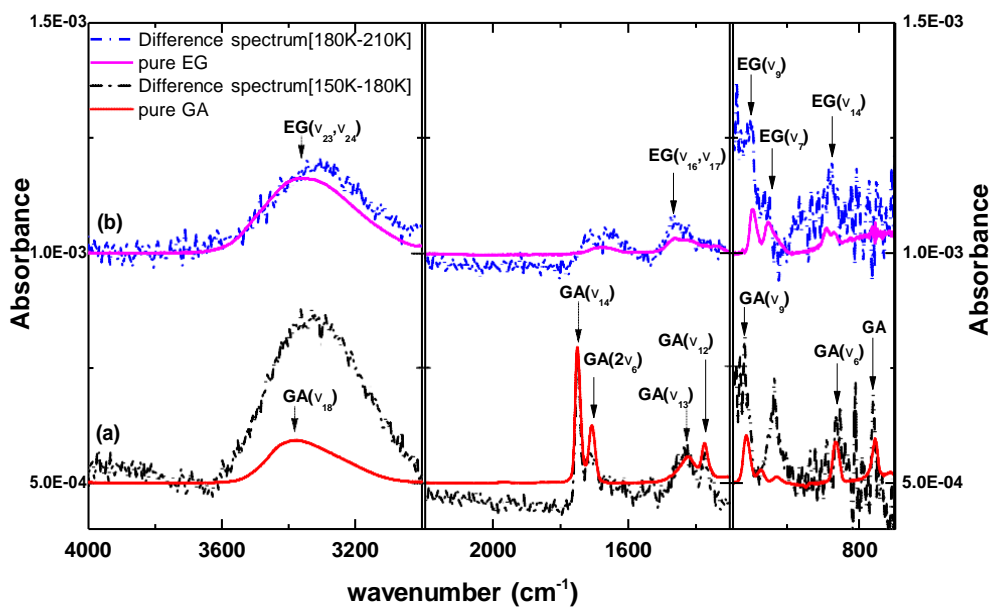

**Figure 3.** RAIR difference spectra (dash-dotted lines) are presented for the experiment of CO:CH$_3$OH = 4:1 ice mixtures with H-atom flux 6.0 × 10$^{12}$ atoms cm$^{-2}$ s$^{-1}$ at 14 K (exp. 1.3) during TPD between 150 and 180 K (a) as well as between 180 and 210 K (b). These are compared to pure GA and EG RAIR spectra (solid lines), respectively. The arrows indicate the IR absorption peaks of the corresponding COMs.

Furthermore, in Fig. 3, the comparison between the difference spectra (dash dotted lines) obtained for chosen temperature ranges during TPD and the spectra of pure GA and EG ice (solid lines) provides additional evidence that both GA, sublimating between 150 and 180 K (a), and EG, sublimating between 180 and 210 K (b), are formed. Two absorption peaks around 1036 and 3320 cm$^{-1}$ are likely due to thermal co-desorption of trapped CH$_3$OH with the aforementioned COMs.

The presence of COMs in the ice upon H-atom addition can be explained through two mechanisms that may take place simultaneously. The first one has been previously discussed by Woods et al. (2012) and experimentally verified by Fedoseev et al. (2015). First, the hydrogenation of CO leads to the formation of HCO intermediate radicals. Their recombination, in turn, yields glyoxal (HC(O)CHO) as a side product. The latter one is subsequently hydrogenated to form GA and EG via

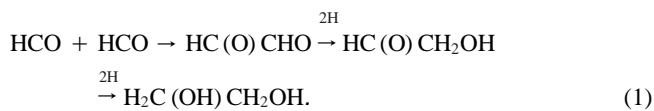

$$\text{HCO} + \text{HCO} \rightarrow \text{HC(O)CHO} \xrightarrow{2H} \text{HC(O)CH}_2\text{OH}$$
$$\xrightarrow{2H} \text{H}_2\text{C(OH)CH}_2\text{OH}. \quad (1)$$

The second mechanism has been described by Chuang et al. (2016). H-atom additions and H-atom-induced abstractions in the CO ↔ H$_2$CO ↔ CH$_3$OH reaction network result in the formation of HCO, CH$_3$O and CH$_2$OH intermediate radicals. Subsequent radical–radical recombination leads to the formation of GA and further hydrogenation results in the formation of EG:

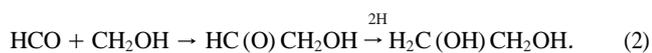

$$\text{HCO} + \text{CH}_2\text{OH} \rightarrow \text{HC(O)CH}_2\text{OH} \xrightarrow{2H} \text{H}_2\text{C(OH)CH}_2\text{OH}. \quad (2)$$

The successive formation of HC(O)CH$_2$OH by recombination of HCO and CH$_2$OH radicals was also reported by Butscher et al. (2015). EG can also be formed by the recombination of two CH$_2$OH radicals. Although not found in the IR spectra due to its low production yield and the limited detection sensitivity of RAIRS, MF is formed through the radical–radical recombination of HCO and

CH$_3$O radicals. Here, only traces of MF can be observed by means of TPD QMS (e.g. exp. 1.3).

### 3.1.1 Overall deposition rate and H-atom/ice ratio dependences

In this paragraph, we focus on the influence of the overall H-atom and CO:CH$_3$OH deposition rates as well as H-atom to CO:CH$_3$OH co-deposition ratios on the formation efficiency of MF, EG and GA. In Fig. 4, zoom-in views of the TPD QMS spectra of CO:CH$_3$OH ice mixtures co-deposited with H atoms are shown for different experimental conditions, i.e. for the same H-atom to CO:CH$_3$OH co-deposition ratio, but for two different overall combined deposition rates (exps. 1 versus 1.2); and for the same number of deposited CO:CH$_3$OH molecules, but for two different H-atom fluxes (exp. 1.2 versus 1.3). The H-atom and CO:CH$_3$OH fluxes in exp. 1 amount to 6.0 × 10$^{12}$ atoms cm$^{-2}$ s$^{-1}$ and 1.2 × 10$^{13}$ molecules cm$^{-2}$ s$^{-1}$ during 60 minutes of co-deposition. In exp. 1.2, these values are 1.0 × 10$^{12}$ atoms cm$^{-2}$ s$^{-1}$ and 2.0 × 10$^{12}$ molecules cm$^{-2}$ s$^{-1}$ during 360 minutes of co-deposition, i.e. the same final fluences of ~2.2 × 10$^{16}$ atoms cm$^{-2}$ and ~4.4 × 10$^{16}$ molecules cm$^{-2}$ are obtained. In exp. 1.3, the same flux of H atoms as in exp. 1 is applied (6.0 × 10$^{12}$ atoms cm$^{-2}$ s$^{-1}$), while the CO:CH$_3$OH flux is kept the same as in exp. 1.2 (2.0 × 10$^{12}$ molecules cm$^{-2}$ s$^{-1}$) for the duration of 360 minutes, resulting in a six times higher fluence of H atoms (~1.3 × 10$^{17}$ atoms cm$^{-2}$) and an unchanged number of deposited CO:CH$_3$OH molecules (~4.4 × 10$^{16}$ molecules cm$^{-2}$) in exps. 1 and 1.2–1.3. The blank experiment (exp. 1.1) is shown for reference. The three left-hand panels in Fig. 4 show the selected TPD QMS spectra and the relevant $m/z$ signals for our three COMs. For MF, the sublimation temperature is around 120 K, and mass scans for the two strongest $m/z$ fragment signals, i.e. CH$_3$O$^+$ (31$m/z$) and CH$_3$OH$^+$ (32$m/z$), as well as the precursor signal HC(O)OCH$_3$$^+$ (60$m/z$) are presented. The desorbing peaks around 160 K assigned to GA are found for three (isomer) mass signals, i.e. CH$_2$OH$^+$ (31$m/z$), CH$_3$OH$^+$ (32$m/z$) and HC(O)CH$_2$OH$^+$





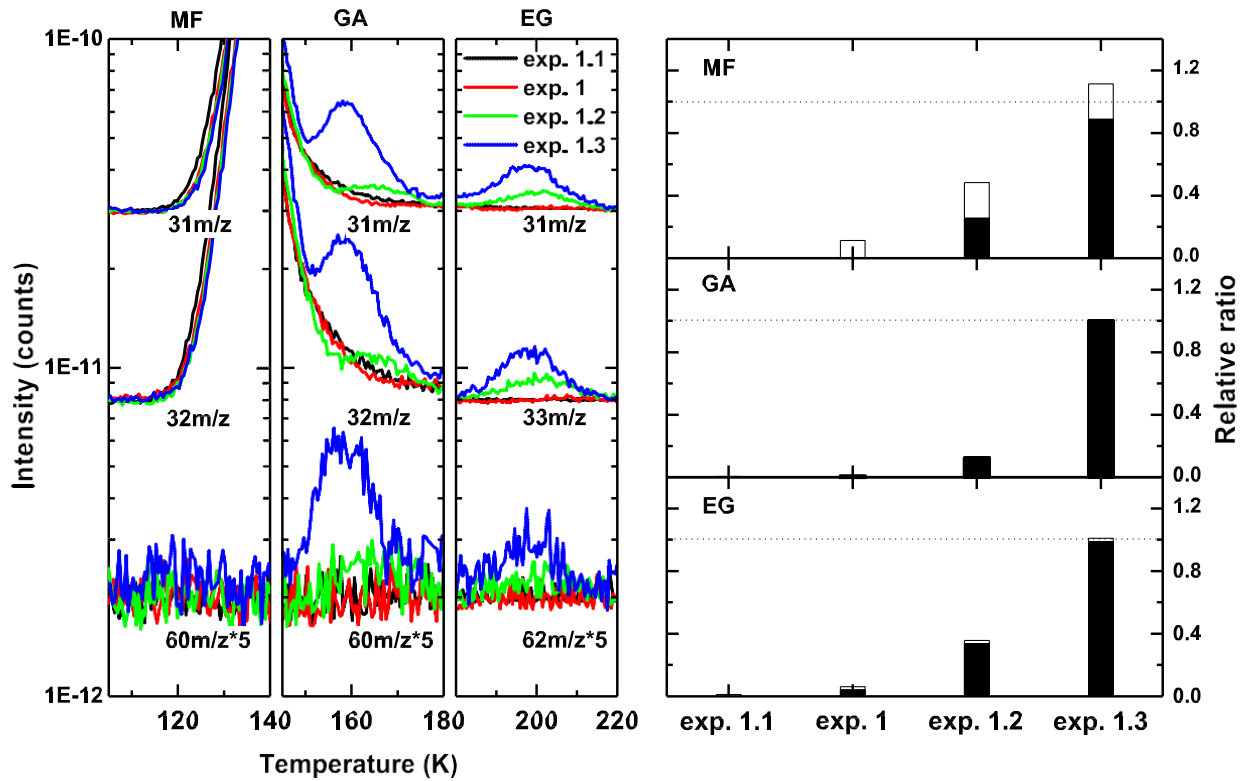

**Figure 4.** Left-hand panels: the TPD QMS spectra obtained after co-deposition of CO:CH$_3$OH = 4:1 ice mixtures with H atoms (exps. 1, 1.2 and 1.3) and blank experiment (exp. 1.1) at 14 K; $m/z$ channels are shifted for clarity. The H-atom fluxes in exps. 1, 1.2 and 1.3 amount to $6.0 \times 10^{12}$, $1.0 \times 10^{12}$ and $6.0 \times 10^{12}$ atoms cm$^{-2}$ s$^{-1}$, while the total H-atom fluences are $2.2 \times 10^{16}$, $2.2 \times 10^{16}$ and $1.3 \times 10^{17}$ atoms cm$^{-2}$, respectively. The total amount of deposited CO and CH$_3$OH molecules is kept constant in all four experiments. Only the relevant $m/z$ numbers and selected desorbing temperatures are shown for MF, GA and EG. Right-hand panels: relative comparison of the integrated intensities of MF (upper panel), GA (middle panel) and EG (lower panel) obtained by TPD QMS and normalized for the total amount of CO, H$_2$CO and CH$_3$OH observed prior to the TPD experiment. Subsequently, each signal is normalized to its maximum yield. The white bars indicate the standard errors derived statistically from an independent set of experiments.

(60$m/z$). For EG, the sublimating peak is around 200 K, and its ionization fragments are CH$_2$OH$^+$ (31$m/z$), CH$_3$OH$_2^+$ (33$m/z$) and the molecular precursor signal H$_2$C(OH)CH$_2$OH$^+$ (62$m/z$).

Besides the qualitative identification of the newly formed COMs through their TPD QMS spectra, also quantitative information can be retrieved by integrating their desorption profiles and normalizing them for the total amount of CO, H$_2$CO and CH$_3$OH, i.e. those species that carry the largest amount of carbon in the ice. The results are shown on the right-hand panels of Fig. 4 and presented as MF, GA and EG, each normalized to its maximum yield.

The blank experiment (CO + CH$_3$OH + H$_2$, exp. 1.1) exhibits, as one expects, no evidence for COM formation. In experiment 1, GA and EG are produced but in much smaller amounts compared to exps. 1.2 and 1.3. The formation follows the mechanisms mentioned before, and the ineffective formation of MF in exps. 1, 1.2 and 1.3 is consistent with previous studies that link this observation to the lack of the intermediate CH$_3$O radicals in the ice bulk, when a CO:CH$_3$OH mixture is co-deposited with H atoms (Chuang et al. 2016). In experiment 1.2, the H-atom and CO:CH$_3$OH fluxes are both reduced by a factor of 6 compared to exp. 1, but the total H-atom and CO:CH$_3$OH fluence in exps. 1 and 1.2 remains the same. It is found that in exp. 1.2 the amount of produced GA and EG is 15 and 7 times higher than in exp. 1, respectively. This is an important experimental finding as the higher production efficiency comes with a lower overall deposition rate. Moreover, traces of MF are found in exp. 1.2. Finally, exp. 1.3 has the same molecule flux as exp. 1.2, but six times higher H-atom flux, i.e. identical to exp. 1.

Thus, exp. 1.3 results in the same number of exposed molecules as in experiment 1 and 1.2 but is performed with six times higher total H-atom fluence. The measured abundances of GA, EG and MF in exp. 1.3 are now 8, 3 and 3 times higher, respectively, in comparison to exp. 1.2 resulting in GA/EG ratios changing from 0.6 to 2.

Experiment 1.2, i.e. with a lower H-atom and CO:CH$_3$OH deposition rate than exp. 1, aims to better mimic the grain surface reactions under dense cloud conditions. The decrease of overall co-deposition flux and the corresponding increase in reaction time reduce the number of barrierless H–H recombination events that fully dominate the surface chemistry when high H-fluxes are applied at short laboratory time-scales. Here, the low accretion rates of H atoms allow them to scan a larger part of an ice surface before meeting another H atom (Fuchs et al. 2009). Thus, a lower H-atom deposition rate enhances the possibility of H-atom addition and abstraction reactions, which, in turn, increases the amount of formed intermediate radicals and COM formation efficiencies. The purpose of exp. 1.3 is to further enhance the number of effective interactions between H atoms and CO:CH$_3$OH molecules by increasing the H-atom flux with a factor of six compared to exp. 1.2 and keeping the CO:CH$_3$OH deposition rate as used in exp. 1.2; this way molecules and radicals will be exposed to multiple interactions with H atoms before the limited H-atom penetration depth in the ice will prevent species to be further (de)hydrogenated (Cuppen et al. 2009; Fuchs et al. 2009). Again differences in the COM composition fractions are observed, as a function of overall deposition rate (comparison between exps. 1 and 1.2) and of H atom/ice ratio (comparison





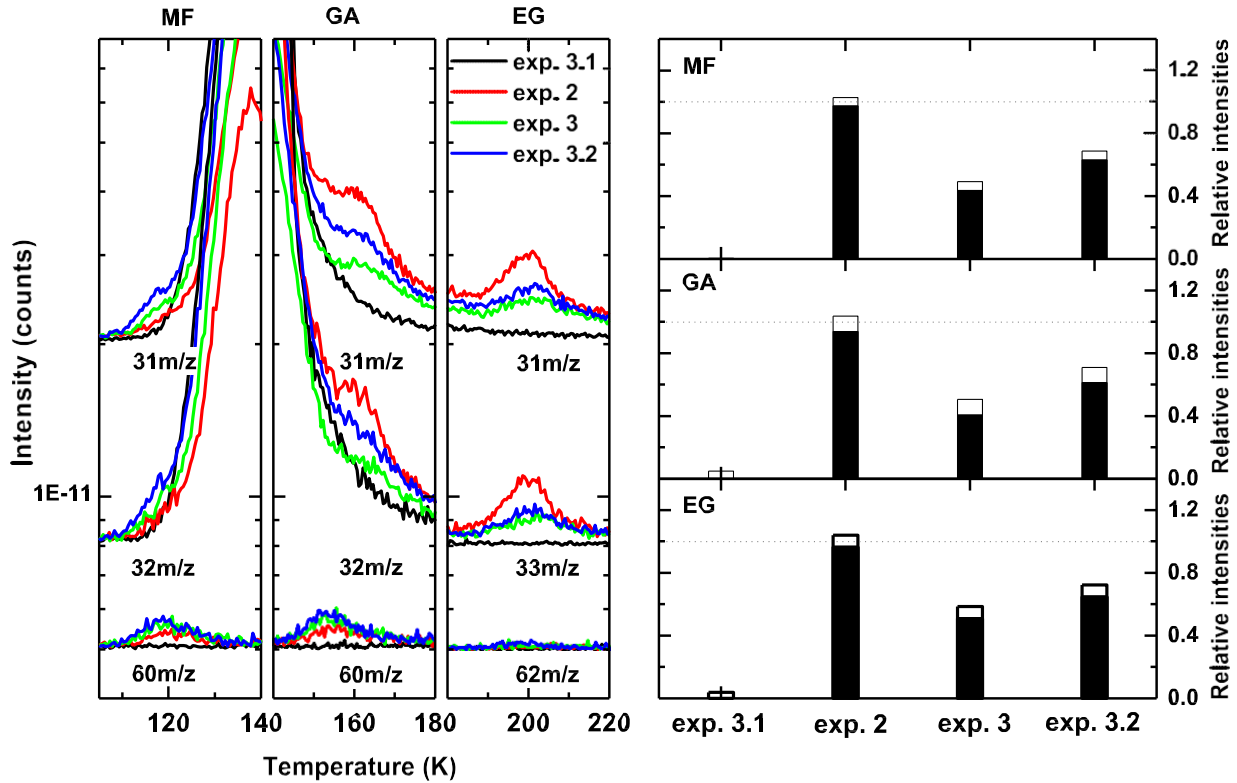

**Figure 5.** Left-hand panels: TPD QMS spectra obtained after co-deposition of CO:CH₃OH ice mixtures in a ratio of 4:1 with UV photons (exp. 2), simultaneous UV photons and H atoms (exp. 3), UV photons and H₂ molecules (exp. 3.2), as well as a blank experiment (exp. 3.1) at 14 K. The $m/z$ channels are presented with offsets for clarity. The UV-photon flux amounts to $4.0 \times 10^{12}$ photons cm⁻²s⁻¹ for exps. 2, 3 and 3.2. Only relevant $m/z$ numbers and selected desorbing temperatures are shown for MF, GA and EG. Right-hand panels: relative comparison of the integrated intensities of MF (upper panel), GA (middle panel) and EG (lower panel) obtained by TPD QMS and normalized for the total amount of carbon-bearing molecules observed prior to TPD. Subsequently, each signal is normalized to its maximum yield. White bars show the standard errors estimated statistically from a set of independent experiments.

between exps. 1.2 and 1.3). The ratio of GA/EG changes from 0.3 to 0.6 to 2 in exps. 1.1, 1.2 and 1.3, respectively. The possible explanation for the enhanced GA/EG ratio is the fact that the formation of GA requires two times less H-atom addition events than the formation of EG (see reactions 1 and 2). These larger amounts of GA with respect to EG follow the increase in the overall COM formation yields. Only traces of MF can be detected in exps. 1.2 and 1.3. This is explained by the lack of CH₃O radicals that are mainly formed through the hydrogenation of H₂CO, a species formed in the ice and not present in the initial sample (Chuang et al. 2016). To sum up, three COMs can be formed upon hydrogenation of CO-rich ices and the relative formation efficiencies depend not only on effective H-atom/(CO:CH₃OH) deposition ratio, but also on the overall accretion rate of the involved species. MF is not abundantly formed in these experiments.

## 3.2 UV photolysis and combined UV and H-atom exposure of CO:CH₃OH ice mixtures

Experiment 2 is performed to study the effect of UV photolysis of CO:CH₃OH ice mixtures on the formation of COMs in the ice. Experiment 3 focusses on the effect of simultaneous hydrogenation and UV irradiation of the same initial ice, while exps. 3.1–4.3 are mainly performed to control that the undissociated H₂ molecules present in the H-atom beam and those recombined on the surface of the ice do not participate in the reaction network. Fig. 5 (three left-hand panels) shows the TPD QMS spectra for exps. 2, 3 and 3.2 (i.e. simultaneous deposition of UV photons and H₂ molecules).

The results for the blank experiment (exp. 3.1) are also shown. In particular, the left-hand panels of Fig. 5 show the relevant $m/z$ signals for MF (31, 32 and 60 $m/z$), GA (31, 32 and 60 $m/z$) and EG (31, 33 and 62 $m/z$) at 120, 160 and 200 K, respectively, that can be interpreted as discussed in Section 3.1. In an identical way, the COM abundances obtained by integrating TPD QMS spectra are normalized for the final amount of the carbon-bearing molecules CO, H₂CO, CH₃OH and CH₄ molecules (right-hand panels of Fig. 5), and MF, GA and EG are each normalized to their maximum yield. The reason for also including CH₄ here, in contrast to exps. 1 and 1.1–1.3, is that only in the UV-photolysis experiments (exps. 2, 3 and 3.1–4.3), methane formation ($0.5$–$1 \times 10^{15}$ molecules cm⁻²) is expected and confirmed by RAIR spectra obtained at the end of the co-deposition. Note that the highest yields for the three COMs under investigation here are found in the pure UV experiment (exp. 2), whereas in the two other experiments (exps. 3 and 3.2) the simultaneous use of UV irradiation and H-atom bombardment substantially reduces the formation rate. In exp. 3.2, the COM yields with respect to those in exp. 2 amount to ∼0.7, which is slightly higher than 0.5–0.6 found in exp. 3. As expected, the blank experiment (CO + CH₃OH + H₂, exp. 3.1) shows no evidence for COM formation.

Given the low gas-phase CH₃OH abundances during deposition we can exclude that any UV-triggered gas-phase processes substantially affect the solid-state reactions that are studied here. Öberg et al. (2009) published the results of an experiment for a pre-deposited CO:CH₃OH = 1:1 ice mixture exposed to UV photons at 20 K, and identified the formation of MF, EG, GA as well as other COMs. The present set of experiments is different, as the irradiation takes place





during the deposition, with a different $CO:CH_3OH = 4:1$ ratio, and at lower temperature (14 K; e.g. exp. 2). The deposited molecules are irradiated by the incoming UV photons with an energy up to 10.2 eV, which is enough to photo-dissociate $CH_3OH$ into radicals, i.e. $CH_2OH$, $CH_3O$, $CH_3$, OH, etc., but not enough to directly photo-dissociate CO molecules. The newly formed intermediate radicals recombine directly to form COMs in the bulk of the ice, resulting in the formation of MF, GA and EG via

$$HCO + CH_3O \rightarrow HC(O)OCH_3 \qquad (3)$$

$$HCO + CH_2OH \rightarrow HC(O)CH_2OH \qquad (4)$$

$$CH_2OH + CH_2OH \rightarrow H_2C(OH)CH_2OH \qquad (5)$$

or other COMs (Öberg et al. 2009).

In dense clouds, the H-atom flux is larger than the UV-photon flux. However, from a comparison between the published $H_2CO$ formation rate values upon UV photolysis of $CH_3OH$ ice and hydrogenation of CO ice (Gerakines et al. 1996; Fuchs et al. 2009; Öberg et al. 2009), it becomes clear that the efficiency of producing radicals by photo-dissociation is (much) larger than by hydrogenation. Moreover, Öberg et al. (2009) discussed the possibility that the presence of H atoms on the grain surface consumes 'frozen-in' radicals, leading to the formation of smaller products before they can grow larger species. In their pre-deposition experiment, this effect is not very significant because the H-atom penetration depth is limited to a few (top) monolayers and H atoms formed upon UV photolysis will mostly leave the surface because of the higher temperature. In the present work, we co-deposit ice constituents with H atoms and $H_2$ molecules during UV-irradiation (exps. 3 and 3.2, respectively). The recombination of H atoms and radicals produced upon photolysis, therefore, will be more prominent, increasing the efficiency of this chemical pathway. For instance, in exp. 3, the abundant H atoms recombine with the photo-induced radicals instead of participating in further hydrogenation of the parent ice mixtures. This results in an enhancement of products, such as $CH_4$ and $H_2O$, as confirmed by RAIR spectra, through the reactions:

$$H + CH_3 \rightarrow CH_4 \qquad (6)$$

$$H + OH \rightarrow H_2O. \qquad (7)$$

and that have been studied before, see e.g. Cuppen et al. (2010).

The results presented in Fig. 5 focus on the formation of MF, GA and EG. These are also the main COM products upon our H-atom addition reactions, but their formation efficiency is influenced as a result of the competition between COM formation (reactions 3–5) and competing hydrogenation reactions (reactions 6–7). In both exps. 3 and 3.2, molecular hydrogen is co-deposited with $CO:CH_3OH$ during UV irradiation, in exp. 3.2 as a main constituent and in exp. 3 mainly as a result of an incomplete dissociation process, which amounts to ∼70 per cent of the hydrogen molecules deposited in exp. 3.2. In previous studies (Fuchs et al. 2009; Lamberts et al. 2013), it was shown that the role of $H_2$ in surface chemical processes is rather limited as the involved activation barriers with $H_2$ cannot be easily overcome for temperatures below 15 K. However, in exp. 3.2, we find a clear COM reduction which is less than that found in exp. 3. To understand the role of $H_2$ in the reaction network, Ar is used instead of $H_2$ in co-deposition with UV photons impacting on $CO:CH_3OH$ in a series of control experiments (exps. 4.1–4.3, details are discussed in the Appendix).

Finally, it is also important to consider the role of UV-induced photo-desorption. A large number of laboratory studies have been performed on the photo-desorption rates of CO and $CH_3OH$ ice (Öberg et al. 2007, 2009; Fayolle et al. 2011; Muñoz-Caro et al. 2010; Chen et al. 2014; Bertin et al. 2016; Paardekooper et al. 2016a; Cruz-Diaz et al. 2016), and even though there exist some discrepancies between the exact values, roughly within an order of 10, this does not prevent the ability to assess the role of photo-desorption in the present experiment. With photo-desorption rates of the order of $1 \times 10^{-2}$ ($1 \times 10^{-3}$) for pure CO ice and less than $1 \times 10^{-4}$ for pure $CH_3OH$ ice, typically for ices with temperatures around 15–20 K, an estimate can be made. Applying these rates to the total UV-photon fluence, which is $1.3 \times 10^{16}$ photons $cm^{-2}$, results in upper limits of photo-desorbed molecules of ∼0.1 (0.01) ML for CO and ∼0.001 ML for $CH_3OH$. This effect will not influence our results, specifically as the photo-desorption rates in mixed ices are expected to further decrease (Bertin et al. 2016).

### 3.3 Summary of laboratory results

Fig. 6 compares the laboratory data taken in this work (left-hand panels) with some selected literature data (right-hand panels) of (i) hydrogenation of ice mixtures including $CO/H_2CO/CH_3OH$ at 15 K by Chuang et al. (2016), (ii) UV irradiation of pre-deposited $CH_3OH$ and $CH_3OH:CO$ ice at 20 K by Öberg et al. (2009), studied by RAIR and QMS techniques and (iii) UV exposure of pre-deposited pure $CH_3OH$ ice at 20 K by Paardekooper et al. (2016a), studied by laser desorption post-ionization mass spectrometry.

For the hydrogenation experiment of $CO:CH_3OH$ ice mixtures presented here (exp. 1.2), the MF/EG, MF/GA and GA/EG at 14 K values are $7 \times 10^{-3}$, $1 \times 10^{-2}$ and 0.6, respectively. These values agree within a factor of 2 with those of similar experiments as reported in Chuang et al. (2016) (i.e. MF/EG = $4 \times 10^{-3}$, MF/GA = $1 \times 10^{-2}$ and GA/EG = 0.4). In the ice mixture experiments containing $H_2CO$ molecules (Chuang et al. 2016), i.e. $CO:H_2CO+H$, and $H_2CO:CH_3OH+H$, the MF/EG and MF/GA values are enhanced by a factor of ∼30, while the GA/EG value remains the same. This shows that, as previously discussed, the MF yields in the final COM compositions are substantially increased if the initial ice mixtures contain $H_2CO$. For the hydrogenation experiments of pure ice, i.e. CO+H and $H_2CO+H$, the final COM compositions are GA dominated over the other COMs, most likely because GA is the first product of the glyoxal (HC(O)HCO) hydrogenation reaction or of the HCO and $CH_2OH$ radical–radical recombination. A similar GA/EG ratio is found in exp. 1.3, where $H_2CO$ is also clearly detected.

The MF/EG, MF/GA and GA/EG 14 K values for pure UV photolysis (exp. 2) amount to 0.4, 0.7 and 0.5, respectively, in reasonable agreement with the values reported in Paardekooper et al. (2016a) (i.e. MF/EG = 0.4, MF/GA = 0.6 and GA/EG = 0.7 derived from their fig. 11). The deviation from the Öberg et al. (2009) values of pure $CH_3OH$ ice ($CO:CH_3OH$ ice mixtures) is much larger, namely, upper limit MF/EG = 0.1, MF/GA = 0.8 (2 for $CO:CH_3OH$ ice mixtures) and upper limit GA/EG = 0.1. However, there exist experimental differences that may explain these different values. In Öberg et al. (2009) pre-deposition is used, but as the selected ices are not very thick, this is not expected to be the determining factor. Instead the much higher UV-photon fluence of $2.4 \times 10^{17}$ photons likely pushes photochemistry beyond the point studied in our study and by Paardekooper et al. (2016a). So, this hints for a clear UV fluence dependence as well. Also the spectral emission characteristics of the H₂ discharge lamp differ from one experimental set-up to another. For instance, a higher $H_2$ flow pressure is used in our lamp settings resulting in a larger broadband $H_2$-emission contribution than just Ly-α (see Ligterink et al. 2015).

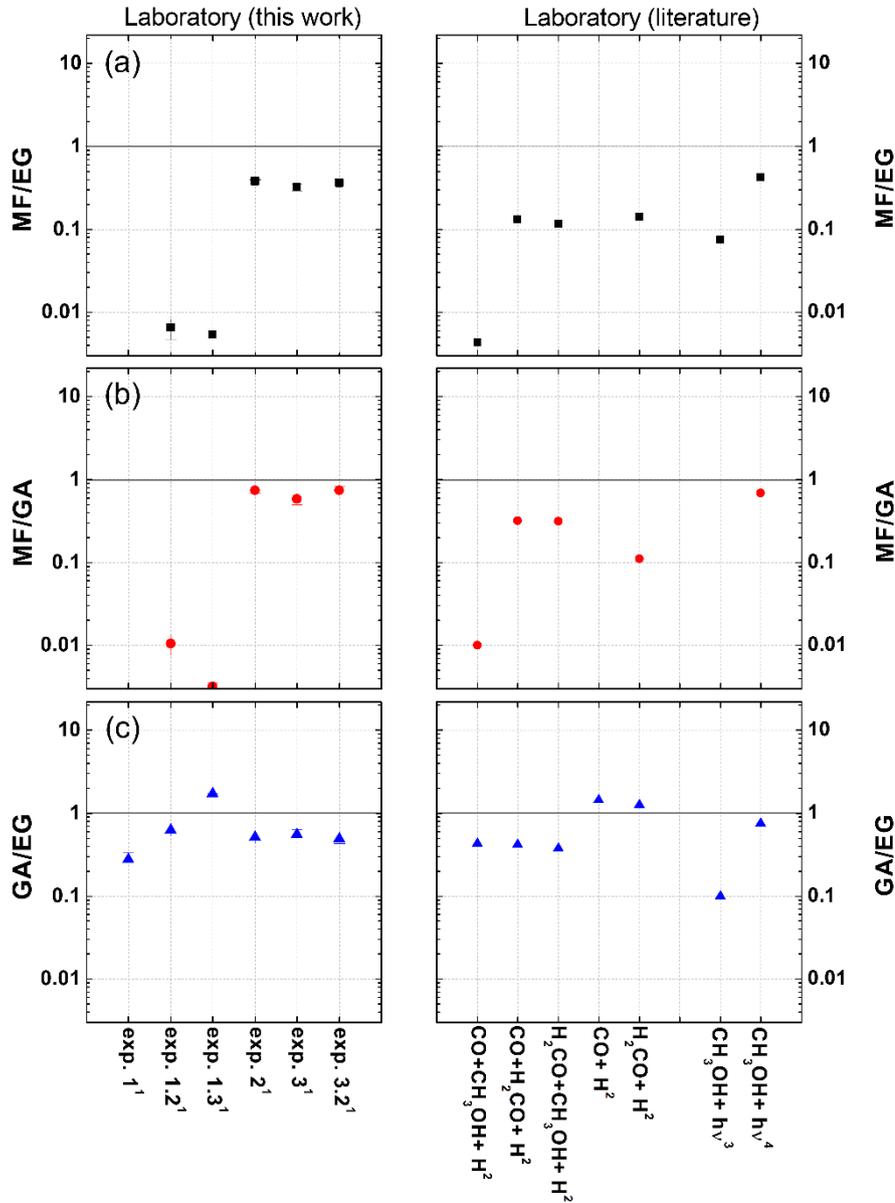

**Figure 6.** An overview of the relative ratios of MF/EG (a), MF/GA (b) and GA/EG (c), derived from laboratory data reported in the literature: (1) this work, (2) Chuang et al. (2016), (3) Öberg et al. (2009) and (4) Paardekooper et al. (2016a).

## 4 ASTROPHYSICAL IMPLICATIONS

It is widely accepted that COMs mainly form in the solid state. This is in agreement with a long list of experiments including those presented here that show how different ('energetic' and 'non-energetic') triggers result in the formation of rather similar species. However, as the underlying formation mechanisms differ, it is to be expected that the final COM abundance ratios in the solid state will vary. In the ISM, MF, EG and GA have only been directly observed in the gas phase. The experiments presented here show clear differences in relative MF, EG and GA ratios for pure hydrogenation, pure UV and combined hydrogenation and UV irradiation experiments.

This is an interesting result and a comparison of the relative ratios of MF/GA, MF/EG and GA/EG offers – in principle – a further constraint to investigate the role of these processes during the chemical evolution of COMs in the ISM. In order to investigate this option, Fig. 7 compares the laboratory data with astronomical values as derived from observations in different astronomical environments.

Observations used in Fig. 7 include several comets (i.e. Hale-Bopp, Lemmon, Lovejoy and 67P/C6) and hot cores of low-mass proto-stars (i.e. IRAS16293-2422B, NGC1333 IRAS2A and NGC1333 IRAS4A) where ices are thermally desorbed. High-mass proto-stars are here excluded because they are generally characterized by a more intense UV field and higher temperatures resulting in further gas-phase chemistry. Comets form at large radii and are thought to harbour the most pristine chemical composition of earlier star-forming regions; in other words, they should preserve the initial pre-stellar ice abundances. Therefore, it is not surprising to find that the MF/EG and GA/EG ratios for the selected comets and low-mass proto-stars are roughly within an order of magnitude with respect to each other (Fig. 7).

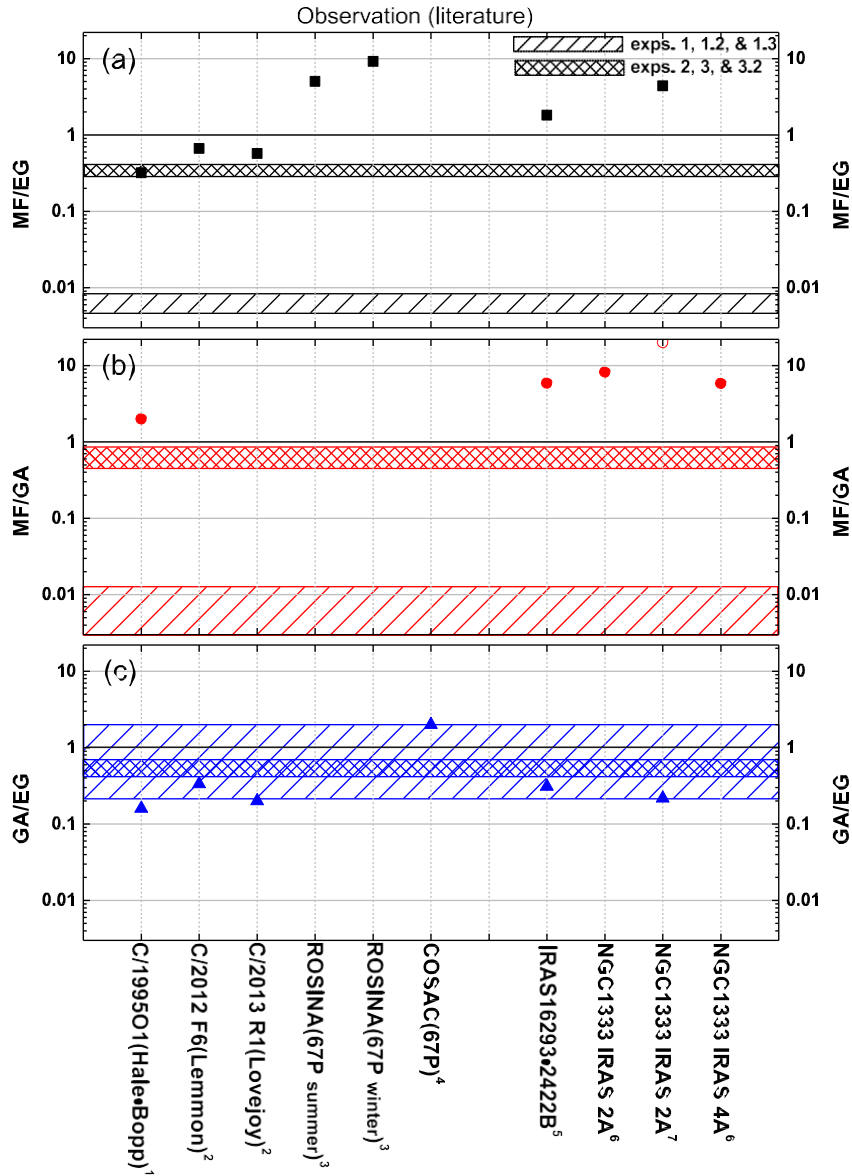

**Figure 7.** An overview of the relative ratios of MF/EG (a), MF/GA (b) and GA/EG (c), derived from astronomical observations reported in the literature. Observational data are taken from (1) Crovisier et al. (2004), (2) Biver et al. (2014), (3) Le Roy et al. (2015), (4) Goesmann et al. (2015), (5) Jørgensen et al. (2016), (6) Taquet et al. (2015) and (7) Coutens et al. (2015). The values of GA/EG and MF/EG for comets are presented as upper limits. In the case of MF/GA this is a lower limit. The shadow areas indicate the ranges of relative ratios taken from exps. 1, 1.2 and 1.3 (hydrogenation only) and exps. 2, 3 and 3.2 (including UV), respectively.

A first comparison of these data to our laboratory results shows that (i) observational MF/EG and MF/GA ratios are up to 3 orders of magnitude higher than found in laboratory hydrogenation experiments of CO:CH₃OH mixtures and up to an order of magnitude higher than experiments that include UV processing; (ii) the observational GA/EG ratio is found in good agreement (i.e. within an order of magnitude) with our ('energetic'/'non-energetic') laboratory data, although hydrogenation experiments seem to reproduce better observations. As stated in previous sections, the lack of formed MF in the hydrogenation experiments comes from the chosen initial ice composition that does not include H₂CO. Results from Chuang et al. (2016) show that the MF/EG and MF/GA ratios obtained from hydrogenation experiments of H₂CO-containing ices are similar to those from the experiments involving UV photolysis shown here. Therefore, COM relative ratios are ice composition dependent in laboratory hydrogenation experiments, more so than for UV-photolysis studies because of the higher amount of radicals formed in the latter process. Simple cold cloud astrochemical models can then be used as a tool to better compare hydrogenation experiments with observations. The astrochemical models described in Cuppen et al. (2009) and Fedoseev et al. (2015) extend the CO hydrogenation scheme as studied in laboratories to dense cloud timescales and conditions. Fedoseev et al. (2015) in their model included the formation of simple ice molecules as well as GA and EG, but not MF. Their results on GA/H₂CO and EG/CH₃OH ratios were concluded to be in good agreement with observations of a solar-type proto-star (IRAS 16293-2422). This is a strong indication that GA and EG are efficiently formed in the solid phase through 'non-energetic' surface reactions in CO-rich ices, while MF formation most likely involves both 'energetic' and 'non-energetic' mechanisms in the solid phase and may even require some additional gas-phase route (Taquet et al. 2015).

A closer look at Fig. 7 shows that ground-based observations, i.e. C/1995 O1-(Hale-Bopp) in Crovisier et al. (2004), C/2012 F6 (Lemmon) and C/2013 R1 (Lovejoy) in Biver et al. (2014), yield upper limit MF/EG ratios of 0.3–0.7 and an upper limit of GA/EG of 0.1–0.2 that are in good agreement as found in the UV-photon irradiation experiments by a factor of ∼3 (exps. 2–3 and 3.2, MF/EG = 0.3–0.4 and GA/EG = 0.5–0.6). The MF/GA = 2 value for comet Hale-Bopp is higher than the experimental UV irradiation results (0.6–0.7), by a factor of ∼3. Although speculative, such a difference may be due to a detection issue; Crovisier et al. (2004) indicated that their upper GA limit may have been underestimated. Another possible explanation is linked to the higher gas-phase reactive nature of GA (Hollis et al. 2001). This would cause GA to be further consumed, decreasing its overall abundance. However, *in situ* detections, like with ROSINA (Le Roy et al. 2015) or COSAC (Goesmann et al. 2015), measuring coma abundances and surface dust compositions for comet 67P provide complementary data. The resulting MF/EG = 5–9 value indicates that MF is more abundant than EG, which is not consistent with other comet observations. Given the different desorption temperatures of MF, EG and other COMs, Le Roy et al. (2015) proposed that the detection of the coma composition depends on the comet surface temperature. This value varies with the comet's orbit and therefore it is not a good reference; MF, for example, desorbs at lower temperatures than EG. In the COSAC passive measurements, the GA/EG value is ∼2 (with an accuracy factor of ∼2). A similar value is only found in our study for the pure hydrogenation result in exp. 1.3 (GA/EG = 2). This is consistent with the picture that comets preserve their pre-stellar COM composition as formed upon hydrogenation reactions in the solid phase. The unambiguous MF detection by COSAC still needs to be confirmed. More recently, cometary ice models, including reactions proposed here, suggest that the abundance of GA is larger than that of MF (Drozdovskaya 2016). The MF/GA ratios in the COM Torus model by Drozdovskaya et al. (2015) consider GA formation routes via H-atom additions.

GA, MF and EG, were detected for the first time toward a solar-mass proto-star by Jørgensen et al. (2012) using ALMA. Observations toward IRAS 2A and 4A also report the presence of these three COMs (Coutens et al. 2015; Taquet et al. 2015). The abundance of MF in these low-mass proto-stars is found to be higher than GA and EG, resulting in ratios of MF/GA and MF/EG larger than 1, i.e. roughly one order of magnitude higher than the experimental results presented here. The aforementioned GA gas-phase destruction rate and the MF enrichment through gas-phase chemistry upon CH₃OH sublimation could contribute to the observed enhanced MF/GA ratio (Balucani et al. 2015; Taquet et al. 2016). Concerning the GA/EG ratio in low-mass proto-stars, Coutens et al. (2015) proposed a value of 0.2 for IRAS 2A, which is of the same order of magnitude as found in exps. 2, 3 and 3.2 (0.5–0.6). The latest updated IRAS 16923B observations (Jørgensen et al. 2016) show a GA/EG ratio of 0.3, consistent with our UV-photon experiments, while still keeping a high abundance of MF.

As stated in Section 1, the comparisons discussed here have to be considered with care. It should be stressed that an abundance ratio in the gas phase does not necessarily reflect that from precursor species in the solid state, especially in cold clouds where thermal desorption, a relatively smooth process, is not possible and other processes are at play. Photo-desorption, for example, can induce substantial fragmentation (Bertin et al. 2016). This will hold for most COMs; however, the ratio of the species that are desorbed without dissociation, in the end, may be representative again. Species may also desorb upon local hot spot formation, due to reactive desorption, i.e. using excess energy generated in a surface reaction, or upon the interaction between the ice grain and a CR. This will clearly influence abundances of species that are formed in first or higher order reaction steps. Moreover, we have focussed here on CO:CH₃OH = 4:1 ices, neglecting H₂CO in the initial mixtures. It has been shown that formaldehyde-enriched ices are more effective in the formation of MF and that other astronomically relevant mixing ratios can lead to different COM formation ratios. Nevertheless, our results strongly support the hypothesis that interstellar solid MF, GA and EG (as well as probably other O-rich COMs) are formed at the same time in a CO ice during and after the catastrophic freeze-out of CO molecules in star-forming regions through a combination of 'energetic' and 'non-energetic' mechanisms.

# 5 CONCLUSIONS

In this work, we present for the first time a quantitative study that compares the influence of pure hydrogenation, pure UV irradiation and both hydrogenation and UV irradiation of CO:CH₃OH = 4:1 ice mixtures at experimental settings and phases typical for the timeframes associated with the CO freeze out stage during the dense cloud collapse. We derive from the laboratory experiments a number of physical chemical dependences:

- Upon hydrogenation of CO-rich ices, MF, GA and EG are formed and their relative formation efficiencies depend on the effective H-atom/(CO:CH₃OH) deposition ratio as well as the used accretion rate of the involved species.

- Whereas GA and EG are formed quite efficiently, MF is produced in much lower abundances in experiments without UV, mainly because the amounts of H₂CO are low. As a consequence, the formation of the CH₃O radical is limited, whereas this has been found to be a primary precursor in the formation of MF.

- Upon UV irradiation, MF, GA and EG are formed (as well as several other COMs), but now MF is produced more efficiently as in this energetic scheme additional reaction pathways become accessible.

- In the combined H-atom and UV-photon exposure study, radicals formed upon photolysis are further hydrogenated; this decreases the overall formation efficiency but does not qualitatively affect the relative abundances.

- The formation ratios for MF:GA:EG for the three different types of experiments and a number of selected settings are summarized in Table 1 and Fig. 1. These roughly range from 0:0.2–0.4:0.8–0.6 for pure hydrogenation to 0.2:0.3:0.5 for pure UV photolysis and combined hydrogenation and UV photolysis.

The astronomical take home message of these findings is that

- The different formation ratios for MF:GA:EG offer a diagnostic potential to derive the formation origin of these species in the solid state in different environments in space (Fig. 7).

- Specifically, the GA/EG ratio observed in comets and in solar-mass proto-stars is consistent with the laboratory values derived for CO-rich ices in the hydrogenation, UV photolysis and combined experiments, and hints at pure solid-state formation pathways.

- Observed ratios involving MF (MF/GA and MF/EG) are not well reproduced in the laboratory experiments; typically we see

less MF formed in the solid state than observed astronomically in the gas phase.

- Future experiments also adding $H_2O$ in the CO:CH$_3$OH mixture may result in MF/GA and MF/EG ratios that are closer to the astronomically data currently available. If not, this may imply that MF in the ISM may also be formed in other ways, e.g. by gas-phase reactions.

## ACKNOWLEDGEMENTS

This research was funded through a VICI grant of NWO (the Netherlands Organization for Scientific Research) and an A-ERC grant 291141 CHEMPLAN. Financial support by NOVA (the Netherlands Research School for Astronomy) and the Royal Netherlands Academy of Arts and Sciences (KNAW) through a professor prize are acknowledged. The described work has had much advantage from collaborations within the framework of the FP7 ITN LASSIE consortium (GA238258). SI acknowledges the Royal Society for financial support. We thank D. M. Paardekooper and V. Taquet for discussions on the laser interference measurements and gas-phase formation of COMs, respectively. We also acknowledge Rachel James for careful proof reading.

## REFERENCES

Arce H. G., Santiago-García J., Jørgensen J. K., Tafalla M., Bachiller R., 2008, ApJ, 681, L21
Bacmann A., Taquet V., Faure A., Kahane C., Ceccarelli C., 2012, A&A, 541, L12
Balucani N., Ceccarelli C., Taquet V., 2015, MNRAS Lett., 449, L16
Belloche A., Müller H. S. P., Menten K. M., Schilke P., Comito C., 2013, A&A, 559, 47
Bertin M. et al., 2016, ApJ, 817, L12
Biver N. et al., 2014, A&A, 566, L5
Blake G. A., Sutton E. C., Masson C. R., Phillips T. G., 1987, ApJ, 315, 621
Boogert A. C. A., Gerakines P. A., Whittet D. C. B., 2015, ARA&A,53, 541
Bottinelli S. et al., 2010, ApJ, 718, 1100
Bull J. N., Harland P. W., 2008, IJMS, 273, 53
Butscher T., Duvernay F., Theule P., Danger G., Carissan Y., Hagebaum-Reignier D., Chiavassa T., 2015, MNRAS, 453, 1587
Caselli P., Ceccarelli C., 2012, A&ARv, 20, 56
Cernicharo J., Marcelino N., Roueff E., Gerin M., Jiménez-Escobar A., Muñoz Caro G. M., 2012, ApJ, 759, L43
Charnley S. B., Tielens A. G. G. M., Millar T. J., 1992, ApJ, 399, L71
Charnley S. B., Rodgers S. D., 2005, in Lis D. C., Blake G. A., Herbst E., eds, Proc. IAU Symp., 231, Astrochemistry, Recent Successes and Current Challenges. Proceedings of the 231st Symposium of the International Astronomical Union held in Pacific Grove. Cambridge Univ. Press, Cambridge.
Chen Y.-J., Chuang K.-J., Muñoz Caro G. M., Nuevo M., Chu C. -C., Yih T.-S., Ip W.-H., Wu C.-Y. R., 2014, ApJ, 781, 15
Chuang K.-J., Fedoseev G., Ioppolo S., van Dishoeck E. F., Linnartz H., 2016, MNRAS, 455, 1702
Codella C., Fontani F., Ceccarelli C., Podio L., Viti S., Bachiller R., Benedettini M., Lefloch B., 2015, MNRAS Lett., 449, L11
Congiu E. et al., 2012, ApJ, 750, L12
Coutens A., Persson M. V., Jørgensen J. K., Wampfler S. F., Lykke J. M., 2015, A&A, 576, A5
Crovisier J., Bockelée-Morvan D., Biver N., Colom P., Despois D., Lis D. C., 2004, A&A, 418, L35
Cruz-Diaz G. A., Martín-Doménech R., Muñoz Caro G. M., Chen Y.-J., 2016, A&A, 592, A68
Cuppen H. M., van Dishoeck E. F., Herbst E., Tielens A. G. G. M., 2009, A&A, 508, 275

Cuppen H. M., Ioppolo S., Romanzin C., Linnartz H., 2010, Phys. Chem. Chem. Phys., 12, 12077
Cuppen H. M., Penteado E. M., Isokoski K., van der Marel N., Linnartz H., 2011, MNRAS, 417, 2809
Drozdovskaya M. N., Walsh C., Visser R., Harsono D., van Dishoeck E. F., 2015, MNRAS, 451, 3836
Drozdovskaya M. N., Walsh C., van Dishoeck E. F., Furuya K., Marboeuf U., Thiabaud A., Harsono D., Visser R., 2016, MNRAS, 462, 977
Fayolle E. C., Bertin M., Romanzin C., Michaut X., Öberg K. I., Linnartz H., Fillion J.-H., 2011, ApJ, 739, L36
Fedoseev G., Cuppen H. M., Ioppolo S., Lamberts T., Linnartz H., 2015, MNRAS, 448, 1288
Fedoseev G., Chuang K.-J., van Dishoeck E. F., Ioppolo S., Linnartz H., 2016, MNRAS, 460, 4297
Fuchs G. W., Cuppen H. M., Ioppolo S., Romanzin C., Bisschop S. E., Andersson S., van Dishoeck E. F., Linnartz H., 2009, A&A, 505, 629
Garrod R. T., Weaver S. L. W., Herbst E., 2008, ApJ, 682, 283
Geppert W. D. et al., 2005, in Lis D. C., Blake G. A., Herbst E., eds, Proc. IAU Symp., 231, Astrochemistry, Recent Successes and Current Challenges. Proceedings of the 231st Symposium of the International Astronomical Union held in Pacific Grove. Cambridge Univ. Press, Cambridge
Gerakines P. A., Schutte W. A., Ehrenfreund P., 1996, A&A, 312, 289
Gibb E. L., Whittet D. C. B., Boogert A. C. A., Tielens A. G. G. M., 2004, ApJS, 151, 35
Goesmann F. et al., 2015, Science, 349, aab0689
Guzmán V. V. et al., 2013, A&A, 560, A73
Henderson B. L., Gudipati M. S., 2015, ApJ, 800, 66
Herbst E., van Dishoeck E. F., 2009, ARA&A, 47, 427
Hidaka H., Watanabe M., Kouchi A., Watanabe N., 2009, ApJ, 702, 291
Hiraoka K., Ohashi N., Kihara Y., Yamamoto K., Sato T., Yamashita A., 1994, J. Chem. Phys. Lett., 229, 408
Hollis J. M., Lovas F. J., Jewell P. R., 2000, ApJ, 540, L107
Hollis J. M., Vogel S. N., Snyder L. E., Jewell P. R., Lovas F. J., 2001, ApJ, 554, L81
Hopkins B. J., Riviere J. C., 1964, Br. J. Appl. Phys., 15, 941
Hudson J. E., Hamilton M. L., Vallance C., Harland P. W., 2003, Phys. Chem. Chem. Phys., 5, 3162
Hudson J. E., Weng Z. F., Vallance C., Harland P. W., 2006, IJMS, 248, 42
Ioppolo S., Fedoseev G., Lamberts T., Romanzin C., Linnartz H., 2013, Rev. Sci. Instrum., 84, 073112
Jiménez-Serra I. et al., 2016, ApJ, 830, L6
Jørgensen J. K., Favre C., Bisschop S. E., Bourke T. L., van Dishoeck E. F., Schmalzl M., 2012, ApJ, 757, L4
Jørgensen J. K. et al., 2016, A&A, 595, 117
Lamberts T., Cuppen H. M., Ioppolo S., Linnartz H., 2013, Phys. Chem. Chem. Phys., 15, 8287
Ligterink N. F. W., Paardekooper D. M., Chuang K.-J., Both M. L., Cruz-Diaz G. A., van Helden J. H., Linnartz H., 2015, A&A, 584, A56
Linnartz H., Ioppolo S., Fedoseev G., 2015, Int. Rev. Phys. Chem., 34, 205
Maity S., Kaiser R. I., Jones B. M., 2015, Phys. Chem. Chem. Phys., 17, 3081
Marcelino N., Cernicharo J., Agúndez M., Roueff E., Gerin M., Martín-Pintado J., Mauersberger R., Thum C., 2007, ApJ, 665, L127
Martín-Pintado J., Rizzo J. R., de Vicente P., Rodríguez-Fernández N. J., Fuente A., 2001, ApJ, 548, L65
McGuire B. A. et al., 2015, ApJ, 812, 76
Mennella V., Baratta G. A., Esposito A., Ferini G., Pendleton Y. J., 2003, ApJ, 587, 727
Minissale M., Dulieu F., Cazaux S., Hocuk S., 2016, A&A, 585, A24
Mathews G. S. et al., 2013, A&A, 557, A132
Modica P., Palumbo M. E., 2010, A&A, 519, A22
Moore M. H., Hudson R. L., 2005, Proceedings of the International Astronomical Union, 1(S231), 247
Mumma M. J., Charnley S. B., 2011, ARA&A, 49, 471
Muñoz Caro G. M., Jiménez-Escobar A., Martín-Gago J. Á., Rogero C., Atienza C., Puertas S., Sobrado J. M., Torres-Redondo J., 2010, A&A, 522, A108
Neill J. L. et al., 2014, ApJ, 789, 8

Öberg K. I., Fuchs G. W., Awad Z., Fraser H. J., Schlemmer S., van Dishoeck E. F., Linnartz H., 2007, ApJ, 662, L23

Öberg K. I., Garrod R. T., van Dishoeck E. F., Linnartz H., 2009, A&A, 504, 891

Öberg K. I., Bottinelli S., Jørgensen J. K., van Dishoeck E. F., 2010, ApJ, 716, 825

Öberg K. I., Boogert A. C. A., Pontoppidan K. M., van den Broek S., van Dishoeck E. F., Bottinelli S., Blake G. A., Evans N. J., 2011, ApJ, 740, 109

Öberg K. I., 2016, Chem. Rev., 116, 9631

Paardekooper D. M., Bossa J.-B., Linnartz H., 2016a, A&A, 592, A67

Paardekooper D. M., Fedoseev G., Riedo A., Linnartz H., 2016b, A&A, 596, A72

Penteado E. M., Boogert A. C. A., Pontoppidan K. M., Ioppolo S., Blake G. A., Cuppen H. M., 2015, MNRAS, 454, 531

Pontoppidan K. M. et al., 2003, A&A, 408, 981

Pontoppidan K. M., van Dishoeck E. F., Dartois E., 2004, A&A, 426, 925

Pontoppidan K. M., 2006, A&A, 453, L47

Prasad S. S., Tarafdar S. P., 1983, ApJ, 267, 603

Pulliam R. L., McGuire B. A., Remijan A. J., 2012, ApJ, 751, 1

Requena-Torres M. A., Martín-Pintado J., Rodríguez-Franco A., Martín S., Rodríguez-Fernández N. J., de Vicente P., 2006, A&A, 455, 971

Requena-Torres M. A., Martín-Pintado J., Martin S., Morris M. R., 2008, ApJ, 672, 352

Rivilla V. M., Beltrán M. T., Cesaroni R., Fontani F., Codella C., Zhang Q., 2016, preprint (arXiv:1608.07491)

Le Roy L. et al., 2015, A&A, 583, A1

Shalabiea O. M., Greenberg J. M., 1994, A&A, 290, 266

Shen C. J., Greenberg J. M., Schutte W. A., van Dishoeck E. F., 2004, A&A, 415, 203

Taquet V., López-Sepulcre A., Ceccarelli C., Neri R., Kahane C., Charnley S. B., 2015, ApJ, 804, 81

Taquet V., Wirström E. S., Charnley S. B., 2016, ApJ, 821, 46

Tercero B. et al., 2015, A&A, 582, L1

Tielens A. G. G. M., Hagen W. W., 1982, A&A, 114, 245

Tielens A. G. G. M., Tokunaga A. T., Geballe T. R., Baas F., 1991, ApJ, 381, 181

Tschersich K. G., 2000, J. Appl. Phys., 87, 2565

Tschersich K. G., Fleischhauer J. P., Schuler H., 2008, J. Appl. Phys., 104, 034908

Vastel C., Ceccarelli C., Lefloch B., Bachiller R., 2014, ApJ, 795, L2

Vasyunin A. I., Herbst E., 2012, ApJ, 762, 86

Vasyunin A. I., Herbst E., 2013, ApJ, 769, 34

Watanabe N., Kouchi A., 2002, ApJ, 571, L173

Watanabe N., Mouri O., Nagaoka A., Chigai T., Kouchi A., Pirronello V., 2007, ApJ, 668, 1001

Woods P. M., Kelly G., Viti S., Slater B., Brown W. A., Puletti F., Burke D. J., Raza Z., 2012, ApJ, 750, 19

Zhitnikov R. A., Dmitriev Y. A., 2002, A&A, 386, 1129

## APPENDIX

The substitution of Ar by $H_2$ results in similar formation yields for MF, EG and GA with relative abundances, 0.71, 0.74 and 0.66, respectively, that are slightly higher (i.e. within their experimental errors) than those in exp. 3.2 (right-hand panels in Fig. 5). Therefore, the most likely explanation for a clear COM reduction in exps.

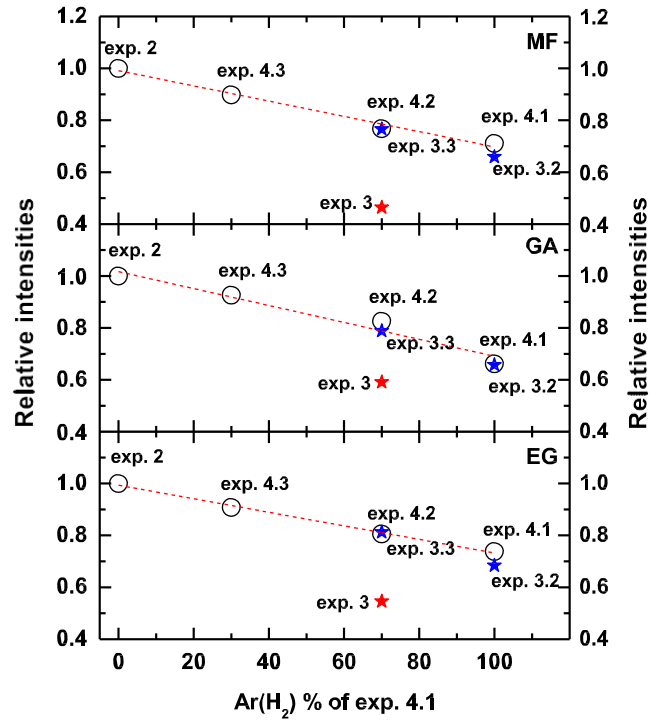

**Figure A1.** Comparison of the integrated intensities of MF (upper panel), GA (middle panel) and EG (lower panel) obtained by TPD QMS for different Ar ($H_2$) flux in exps. 2, 3, 3.2 and 3.3 and for the controlled experiments of CO + $CH_3OH$ + Ar + UV (exps. 4.1–4.3). The integrated intensities are normalized for the total amount of carbon-bearing molecules observed prior to TPD and then each of them is normalized to their maximum yield.

3.2 and 4.1 is that the presence of $H_2$ (or Ar) causes physical segregation of the reactants reducing the recombination probability of HCO, $CH_2OH$ and $CH_3O$ radicals to form complex molecules. Furthermore, the segregation effect seen when using Ar instead of $H_2$ is confirmed by varying the Ar deposition rate, i.e. 100, 70 and 30 per cent of the Ar flux used in exp. 4.1, respectively (see Fig. A1; exps. 4.1–4.3). As shown in Fig. A1, the higher Ar flux gives less COM production efficiency. In exps. 4.2 and 3.3, the chosen Ar and $H_2$ flux is the same as the flux of incomplete dissociated $H_2$ in exp. 3. The difference in COM yields shown in Fig. A1 between exps. 3 and 4.2 as well as 3.3 is purely due to the H-radical recombination forming simple species through reactions (6–7). Although less likely, minor differences between exps. 3.2 and 4.1 can be due to H atoms produced by the photo-dissociation of $H_2$. However, in the gas phase, the direct photo-dissociation of $H_2$ does not occur for photon energies below 11.27 eV. This makes the photo-dissociation of $H_2$ an unlikely scenario under our experimental conditions.

This paper has been typeset from a TEX/LATEX file prepared by the author.